# PHASE BEHAVIOR OF MELTS OF DIBLOCK-COPOLYMERS WITH ONE CHARGED BLOCK

Alexey A. Gavrilov[§*], Alexander V. Chertovich[§,∥] and Igor I. Potemkin[§,≠,#]

[§] Physics Department, Lomonosov Moscow State University, Moscow 119991, Russian Federation
[∥] Semenov Institute of Chemical Physics, Moscow 119991, Russian Federation
[≠] DWI - Leibniz Institute for Interactive Materials, Aachen 52056, Germany
[#] National Research South Ural State University, Chelyabinsk 454080, Russian Federation

*e-mail: gavrilov@polly.phys.msu.ru

ABSTRACT

In this work we investigated the phase behavior of melts of block-copolymers with one charged block by means of dissipative particle dynamics with explicit electrostatic interactions. We assumed that the blocks are fully compatible, i.e. all the Flory-Huggins $\chi$ parameters were equal to 0. We showed that the charge correlation attraction solely can cause microphase separation with long-range order; a phase diagram was constructed by varying the volume fraction of the uncharged block and the electrostatic interaction parameter $\lambda$. The obtained phase diagram was compared to the phase diagram of corresponding neutral diblock-copolymers with non-zero $\chi$-parameter between the beads of different blocks; one block of such copolymers consisted of "dumb-bell" monomer units to mimick the increase of the volume of each monomer due to the presense of counterions. Surprisingly, the differences between these phase diagrams are rather subtle; the same phases in the same order are observed, and the positions of the ODT points are similar if the $\lambda$-parameter is considered as an "effective" $\chi$- parameter. Next, we studied the position of the ODT for lamellar structure depending on the chain length $N$. It turned out that while for the uncharged diblock-copolymer the product $\chi^{cr} N$ was almost independent of $N$, for the diblock-copolymers with one charged block we observed a significant increase in $\lambda^{cr} N$ upon increasing $N$. It can be attributed to the fact that the counterion entropy prevents the formation of ordered structures, and its influence is more pronounced for longer chains since they undergo the transition to ordered structures at smaller values of $\lambda$, when the electrostatic energy becomes comparable to $k_b T$. This was supported by studying the ODT in diblock-copolymers with charged blocks and counterions cross-linked to the charged monomer units (thus forming dipole monomers). The ODT for such systems was observed at significantly lower values of $\lambda$ with the difference being more pronounced at longer chain lengths $N$. The diffusion of counterions in the obtained ordered structures was studied and compared to the case of a system with the same number of charged groups but homogeneous structure; the diffusion coefficient in a direction in the lamellar plane was found to be higher than in any direction in homogeneous structure.

**Introduction**

Microphase separated polymer systems have attracted great attention due to their wide range of possible applications. Despite the significant success in studying the phase behavior of uncharged copolymers, there is no comprehensive description of the problem of microphase separation in polyelectrolytes (polymers containing charged groups). The presence of charged groups and, as a result, the effects related to that (such as correlation attraction, ion association processes, counterion entropy etc.) significantly alters the system behavior. Investigation of the microphase separation in polyelectrolytes is of topical interest because nanostructured ion-conducting membranes have promising applications in a variety of energy storage and conversion devises such as lithium-ion batteries, fuel cells and others[1–3]. Moreover, there are other possible applications of such nanostructures systems in nanoelectronics, nanophotonics, as membranes for separation etc. Microphase separated nanostructured systems have a number advantages compared to homogeneous ones: different phases can have different properties, for instance, one can make the material mechanically stable and tough, while the other can have good ion conductivity. Therefore, understanding the principles of the formation of phases with different topologies in polyelectrolytes will allow creating new materials with significantly improved properties for the aforementioned applications.

During the last years the problem of microphase separation in polyelectrolytes has attracted increasing attention of the researchers. In the majority of works such systems are studied experimentally. As a good example one can refer to the review [4] in which the influence of electrostatic interactions on the phase behavior of copolymers with charged and neutral blocks is discussed. In particular, it was shown[5] that for sulfonated poly(styrene)-*b*-poly(methylbutylene) diblock-copolymer the degree of sulfonation of the charged block plays an important role in determining the system morphology. Varying the sulfonation degree between 0 and 44.7%, the morphologies were found to change from disordered to gyroid to lamellae to hexagonally perforated lamellae (HPL). However, it was noted that no morphologies other than the ones present on the phase diagram of uncharged diblock-copolymer (for example, see [6]) were found. An "unconventional" morphology was found in work[7]: inverted cylinders were observed for sulfonated poly-(styrene)-b-poly(isoprene) with the sulfonated polystyrene minor component forming the matrix. The authors attributed such behavior to the counterions entropy; however, the inverted cylinders were formed only immediately after the film formation, and thermal annealing destroyed the long-range order. It is apparent that there is a need for a significant amount of additional investigations in order to understand the physical principles of formation of microphase separation in such systems[4].

One of the most interesting and dynamically developing areas of polyelectrolytes studies is investigation of the properties of poly(ionic liquids). In short, polyionic liquids are a special type of polyelectrolytes which carry an IL species in each of the repeating units[8]. Such architecture allows one to combine the advantages of polymers and ionic liquids in one material. For instance, the "classical" polymer electrolytes could not play the role of solid ion conductors: due to the extensive ion-pairing only materials with very low bulk conductivity are usually obtained in absence of a polar solvent. A good example of such behavior is a well-known material Nafion, which conducts only swollen in a significant amount of water. This problem is not present for poly (ionic liquids), as they demonstrate high ionic conductivity even in the "dry" state (without solvent), which significantly improves the mechanical properties of such conducting materials. Moreover, there are a lot of types of anions and cations which can be used, and the resulting polymer material can conduct either cations or anions depending on which ion is linked to the polymer. All these circumstances allow one to finely tune the properties of the system. One of the most comprehensive investigations of the phase behavior of block copolymers with one block being a poly(ionic liquid) is the work [9]. In that work perfectly ordered structures of various types were obtained for different fractions of the polyelectrolyte block, including lamellar, cylindrical and spherical. In the work [10] the microphase separated systems formed by diblock-copolymers with poly(ionic liquid) block were shown to have higher conductivity compared to the ones with weak separation; the systems that tend to form bicontinuous structures demonstrated the best conductivity among all others. The influence of the presence of microphase separation on conductivity was also studied in the work[11]; it was found that the hydroxide conductivity is higher in microphase-separated diblock-copolymers with one charged block than in analogous conducting homopolymer (and much higher compared to random copolymers comprised of the same monomers), which was explained by the presence of conducting channels in the microphase-separated systems. More complex chain architectures are also being studied: in a recent paper [12] the combination of properties of triblock copolymers with charged side blocks was found to be better than that described in the literature for other types of electrolytes.

As for the theoretical and simulation works, there is still no more or less systematic description of the microphase separation in polyelectrolytes, which is especially apparent if one considers the phase behavior of neutral copolymers that is studied very well even for different polymer chain architectures. Moreover, due to the presence of several types of interactions and length scales it is a complicated task to create a consistent theoretical description of such systems. One of the most recent and complete investigation of the microphsase separation in melts of diblock-copolymers with one charged block was carried out in the papers[13,14]. In those works, a combination of the self-consistent field theory and the liquid state theory accounting for the local charge organisation was used; the latter play an important, if not the main, role in the self-organization of polyelectrolytes. It was shown that the presence of charged monomer units and differences in the dielectric permittivities between phases formed by different blocks significantly change the phase diagram, making it assymmetrical, and allowing one to obtain ordered structures even at zero Flory-Huggins parameter. Furthermore, it was shown that the region of ordered stucrures dramatically shrinks upon increasing the fraction of charged units higher than 15%, which is rather interesting: it seems that an increase in the fraction of charged units should facilitate the microphase separation because of the intensification of the charge correlation attraction. Therefore, there is a need for a detalied investigation of the system with particle methods, in which the correlation attraction as well as fluctuations (which play in important role in the microphase separation phenomenon) are taken into account automatically.

In this work we addressed the question of microphase separation in melts of diblock-copolymers with one charged block. We focused on the effect of the presence of charged groups and mobile counterions and therefore investigated the case of completely compatible blocks (Flory-Huggins parameter $\chi=0$). The behavior of such systems was compared to that of corresponding uncharged diblock-copolymers as well as copolymers with cross-linked counterions. Finally, we studied the influence of the ordering in the system on the counterion diffusion.

**Method and model**

In this work we used dissipative particle dynamics with explicit electrostatic interactions as the simulation method. First we give a brief description of the standard without electrostatic interactions. Dissipative particle dynamics (DPD) is a version of the coarse-grained molecular dynamics adapted to polymers and mapped onto the classical lattice Flory–Huggins theory [15–18]. It is a well-known method which has been used to simulate properties of a wide range of polymeric systems, such as single chains in solutions, microphase separation in polymer melts and networks. In short, macromolecules are represented in terms of the bead-and-spring model (each coarse-grained bead usually represents a group of atoms), with beads interacting by a conservative force (repulsion) $\boldsymbol{F}_{ij}^{c}$, a bond stretching force (only for connected beads) $\boldsymbol{F}_{ij}^{b}$, a dissipative force (friction) $\boldsymbol{F}_{ij}^{d}$, and a random force (heat generator) $\boldsymbol{F}_{ij}^{r}$. The total force is given by:

$$\boldsymbol{F}_{i} = \sum_{i \neq j} \left( \boldsymbol{F}_{ij}^{c} + \boldsymbol{F}_{ij}^{b} + \boldsymbol{F}_{ij}^{d} + \boldsymbol{F}_{ij}^{r} \right) \tag{1}$$

The soft core repulsion between *i*- and *j*-th beads is equal to:

$$\boldsymbol{F}_{ij}^c = \begin{cases} a_{\alpha\beta}(1 - r_{ij}/R_c)\boldsymbol{r}_{ij}/r_{ij}, & r_{ij} \leq R_c \\ 0, & r_{ij} > R_c \end{cases}, \quad (2)$$

where $\boldsymbol{r}_{ij}$ is the vector between $i$-th and $j$-th bead, $a_{\alpha\beta}$ is the repulsion parameter if the particle $i$ has the type $\alpha$ and the particle $j$ has the type $\beta$ and $R_c$ is the cutoff distance. $R_c$ is basically a free parameter depending on the volume of real atoms each bead represents [18]; $R_c$ is usually taken as the length scale, i.e. $R_c=1$.

If two beads ($i$ and $j$) are connected by a bond, there is also a simple spring force acting on them:

$$\boldsymbol{F}_{ij}^b = -K(r_{ij} - l_0)\frac{\boldsymbol{r}_{ij}}{r_{ij}}, \quad (3)$$

where $K$ is the bond stiffness and $l_0$ is the equilibrium bond length.

We do not give here a more detailed description of the standard DPD model (without electrostatic interactions); it can be found elsewhere [18].

In order to take into account the electrostatic interactions, we use the method described in the work [19]. The electrostatic force between two charged beads is calculated using the following expression:

$$\boldsymbol{F}_{ij}^e = \frac{q_i q_j}{4\pi\varepsilon\varepsilon_0}\begin{cases} \frac{\boldsymbol{r}_{ij}}{r_{ij}^3}\sin^6\left(\frac{2\pi r_{ij}}{4D}\right), & r_{ij} < D \\ \frac{\boldsymbol{r}_{ij}}{r_{ij}^3}, & r_{ij} \geq D \end{cases},$$

where $D$ is the damping distance. This approach allows one to prevent overlapping of oppositely charged species while keeping the exact form of the Coulomb potential at distances larger than D; the parameter D is essential the effective bead size and the electostatic interactions at smaller distances are not important for the system behavior. We used $D=0.65$ which was shown [19] to be a good choice for the number density of 3, which was used in our work.

The melts (no solvent was added) of diblock-copolymers with one charged block with the backbone length of N=24 were considered; each monomer unit of the charged block carried a charge of $+e$, while the counterions had a charge of $-e$ (i.e. the number of counterions was equal to the number of charged monomer units). Since in the classical DPD all the beads have the same size, this is a model of a diblock-copolymer with one block being poly(ionic liquid) with a large free ion. If not mentioned, all the Flory-Huggins interaction parameters in the system were equal to 0. The strength of the electrostatic interactions was characterized by a dimensionless parameter $\lambda = \frac{l_b}{R_c}$, where $l_b$ is the Bjerrum length in the system; it was varied from 1 to 20. If we consider parameterization similar to that proposed by Groot [20] with $R_c \approx 0.7$nm, then $\lambda=1$ would correspond to a polar medium with ε≈80 at room temperature, while λ=20 – to a medium with ε≈4 at room temperature.

The behavior of copolymers with one charged block was compared to that of corresponding uncharged diblock-copolymers. The incompatibility between the beads constituting different blocks was expressed in terms of the Flory-Huggins parameter χ. Using the approach described in the work [18], we found that the relation between the value $\Delta a = a_{\alpha\beta} - a_{\alpha\alpha}$ ($a_{\alpha\beta}$ is the interaction parameter between particles of the types α and β, α ≠β) and χ for the used parameters (interaction parameter $a_{\alpha\alpha}$=50, bond stiffness $K$=4.0) in the case of homopolymers has the following form: $\chi_{\alpha\beta}$=0.253±0.004$\Delta a_{\alpha\beta}$; this expression was used in what follows.

## Results and discussions

First of all, we studied the phase diagram for the diblock-copolymer polyelectrolyte chains with charges located directly on the backbone. The resulting phase diagram is presented in Fig.1, left; it was calculated in the coordinates (φ- λ), where φ is the volume fraction of the uncharged block in the system.

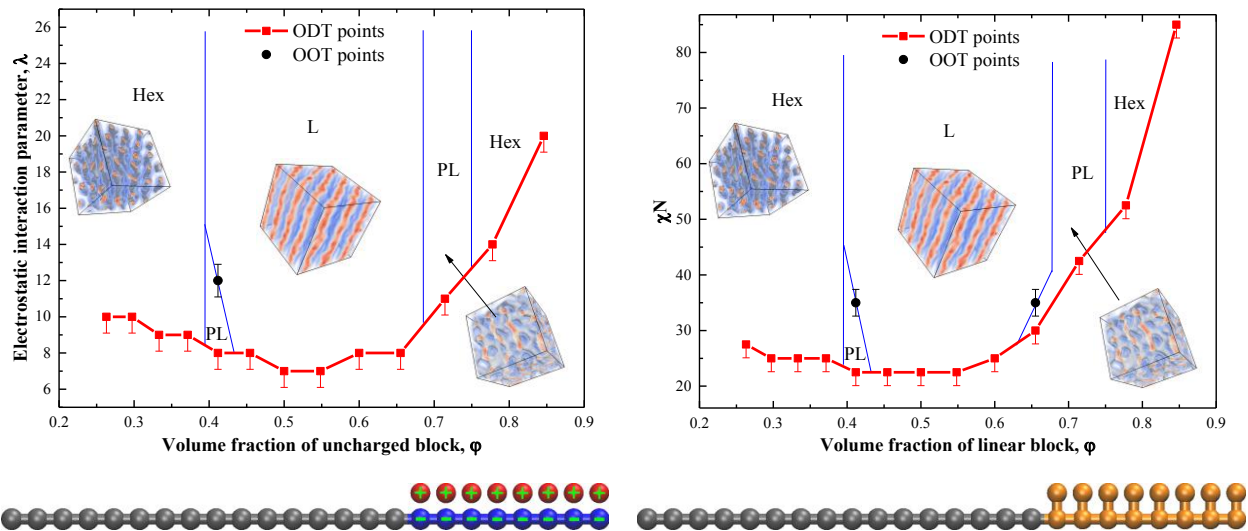

Fig.1 Phase diagrams for melts of diblock-copolymers with one charged block (left) and corresponding uncharged diblock-copolymers (right); L - lamellae, PL – hexagonally perforated lamellae, Hex – hexagonally packed cylinders. The lines show the expected positions of the boundaries between the phases and are given as a guide for the eye. The studied chain architectures at φ=0.5 are shown below each diagram.

The diagram is rather similar to the conventional phase diagrams of uncharged diblock copolymers [6,21,22] in terms of the observed phases and their order; it is pronouncedly asymmetric, however, due to the different architecture of the blocks. The uncharged domains are formed by linear blocks, whereas in the domains containing charges the counterions effectively increase the volume of each monomer unit. The phase diagram was obtained at $\chi=0$ for all interactions, i.e. the observed microphase separation occurs solely due to the charge-correlation attraction [23]; such attraction is believed to be the reason for the collapse of a single polyelectrolyte chain on non-polar solvent [24,25]. Let us know compare the obtained phase diagram to the phase diagram of the corresponding neutral diblock-copolymer where the behavior is governed by the $\chi$-parameter. In some sense, the polyelectrolyte block and its counterions can be considered as backbone with one additional bead attached to each monomer unit (such chain architecture is depicted in Fig.1). The phase diagram was calculated in the coordinates ($\varphi$- $\chi_{AB}$ N), where N was taken equal to the backbone length of 24 and $\chi_{AB}$ is the Flory-Huggins parameter between the beads of different blocks; it is presented in Fig.1, right. Surprisingly, the diagrams for the charged (Fig.1, left) and the corresponding neutral (Fig.1, right) copolymers are almost identical in those coordinates. The observed differences are rather subtle and can be attributed to the discrete variation of the interaction parameters ($\lambda$ and $\chi_{AB}$) and the subsequent error in the order-disorder transition (ODT) position as well as the presence of the counterion entropy in the charged diblock-copolymer melt. The dimensionless parameters $\lambda$ and $\chi_{AB}$ seem to play the same role in the microphase separation phenomenon, which is logical: they are both inversely proportional to $k_bT$. Therefore, the driving force for the microphase separation in the studied diblock-copolymers with one charged block is completely different from that in "classical" diblock-copolymers, but the phase behavior seems to be almost independent of the exact nature of it.

But are the two systems being compared, polyelectrolyte and neutral, are equivalent when it comes to microphase separation? In order to test that, we calculated the dependence of the structure period on the interaction parameters. To that end, we considered the system with $\varphi=0.5$ (i.e. the lamellar-forming copolymer); the box size was increased to $80^3$ to reduce the effect of the finite system size. In order to determine the period of the lamellar structure, we calculated the static structure factor using the standard approach [21]; the position of the first peak gives us the structure period. The obtained dependences are presented in Fig.2.

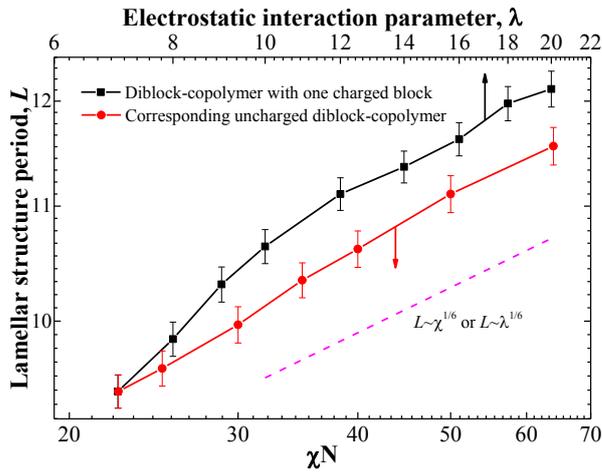

Fig.2 The period of the lamellar structure for $\varphi=0.5$ depending on the interaction parameter: $\lambda$ for the diblock-copolymers with a charged block and $\chi_{AB}N$ for the corresponding neutral diblock-copolymers. The ranges of the X-axes are chosen so that they show the same relative increase of the interaction parameter.

We can see that for the uncharged copolymer, even though one of its blocks has additional beads attached to the backbone, the theoretical dependence for "classical" diblock-copolymers [26] of $L\sim\chi^{1/6}$ (with $L$ being the structure period) is reproduced. The same scaling $L\sim\lambda^{1/6}$ is observed for the diblock-copolymer with one charged block at $\lambda \gtrsim 10$; however, at lower values of $\lambda$, closer to the transition point, the changes in the lamellar period have a stronger dependence on $\lambda$. We suppose that this is due to the counterion entropy: while the system behavior at large $\lambda$ is dominated by the electrostatic energy (it is much larger than $k_bT$ at the length scales of one monomer unit), at smaller $\lambda$ the counterion entropy plays an important role and prevents the formation of ordered structures.

In order to test this hypothesis, we studied the position of the ODT point for different chain lengths for both types of block-copolymers at $\varphi=0.5$; the obtained dependencies are shown in Fig.3.

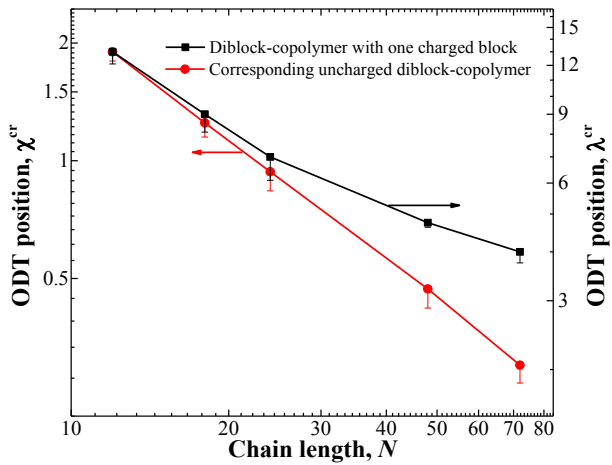

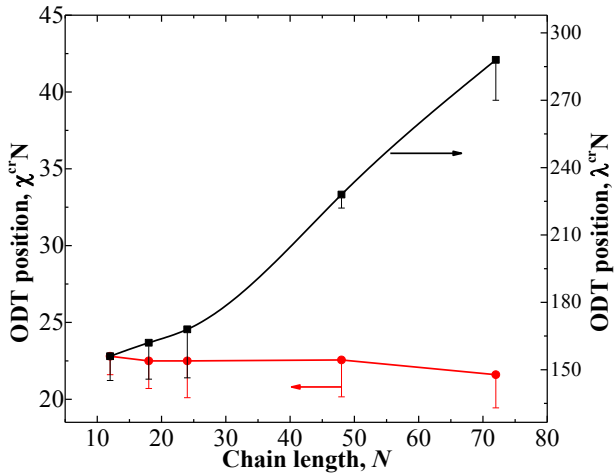

Fig. 3 Positions of the ODT for diblock-copolymers with one charged block and corresponding uncharged diblock-copolymer at $\varphi=0.5$ depending on the chain length. The top graph shows $\lambda^{cr}$ (for diblock-colopymers with one charged block) and $\chi_{AB}^{cr}$ (for corresponding uncharged diblock-colopymers), while the bottom shows the products $\lambda^{cr}N$ and $\chi_{AB}^{cr}N$. The ranges of the Y-axes are chosen so that they show the same relative increase of the interaction parameter.

The dependence of $\chi_{AB}^{cr}$ on $N$ for the uncharged diblock-copolymer is linear (so the product $\chi_{AB}^{cr}N$ is roughly constant), as it was expected; while the presence of fluctuations leads to a shift in the position ODT point depending on the chain length [27], the studied range of chain length is small enough so that the differences are within the error. A very different behavior is observed for the diblock-copolymer with one charged block: the product $\lambda^{cr}N$ significantly increases with the increase in N. The counterion entropy significantly influences the ODT for longer chains since the transition point for them shifts towards smaller values of $\lambda$. We therefore can speculate that in the limit of very long chains the transition point $\lambda^{cr}$ would be approaching some finite value of $\lambda$ upon increasing N since the counterion entropy dominates the system behavior at small values of $\lambda$.

Another way to show the influence of the counterion entropy is to remove it from the system altogether and investigate the differences. To that end, we studied the position of the ODT point for diblock-copolymers with one charged block in which the counterions were cross-linked to the charged monomer units (i.e. forming dipoles). While the structure of the ionic domain could be somewhat different from that observed in the system with free counterions, the general physical reasons behind the microphase separation are the same. The case with $\varphi=0.5$ was investigated; we found that for the chain length of $N=24$ the position of the ODT shifted from $\lambda^{cr}=7\pm1$ to $5\pm1$, i.e. the value becomes ~1.4 times smaller, while for the chains with $N=72$ the difference is much more pronounced: from $\lambda^{cr}$=4±0.5 to 1.6±0.2, i.e. the value becomes ~2.5 times smaller. These data support our hypothesis about the influence of the counterion entropy once again: it is more significant when the microphase separation is observed at small $\lambda$, i.e. for long chains.

As it was mentioned in the Introduction, copolymers with polyelectrolyte blocks are often considered as new-generation materials for ion-exchanging membranes. For such applications, the ion conductivity is a crucial characteristic. Ion conductivity is related to the diffusion of free ions; therefore, we studied the influence of the presence of ordered structures on the ion diffusion. To that end, we compared the mean-squared displacement (MSD) of the counterions in lamellar structures at φ=0.6 to that in melts of copolymers in which the charged groups were distributed evenly, but their number as well as the chain length was the same. The latter copolymers did not form any ordered structures even at the highest studied $\lambda$; the results are presented in Fig.4. The mean-squared displacement for the lamellar structures was calculated along one axis in the lamellar plane (the diffusion perpendicular to the lamellae is limited by the domain size); for the copolymers with evenly distributed charged groups the diffusion is isotropic, so any direction can be chosen.

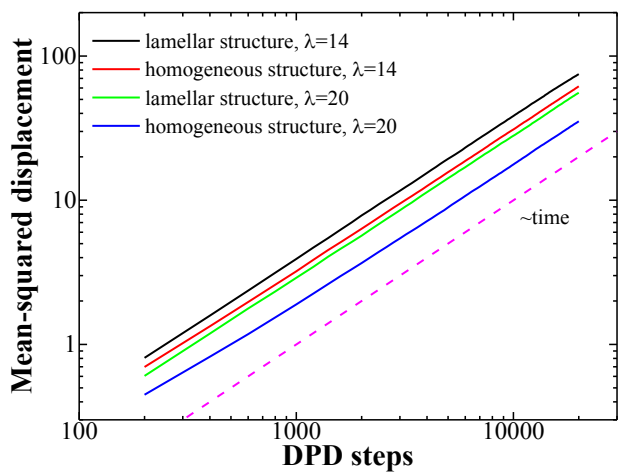

Fig.4 Mean-squared displacement of counterions at different values of $\lambda$ (14 and 20) obtained for lamellar structures (diblock-copolymers) and homogeneous structures (copolymers with evenly distributed charged groups). The diffusion coefficient is reflected by the position of the curve: the higher the curve, the higher the diffusion coefficient.

First of all, we can see that in all studied cases the MSD shows a linear relationship to time, i.e. the counterions undergo normal diffusion. No stable ion pairs are formed, and the counterions are "shared" between all the charged monomers (despite rather high values of $\lambda$) like electrons in metals. Similar behavior was observed in globules formed by a single polyelectrolyte chain [25,28], in which the reason for the formation of the collapsed state is also the charge-correlation attraction. Second, we see that for both studied values of λ the diffusion in the ordered structures is faster; the difference is more pronounced at the higher λ-value.

Due to the homogeneous structure of the melts of copolymers with evenly spaced charged groups, the movement of the counterions is slowed down by the presence of neutral regions in the system; in some sense, there is no direct pathway formed by charged monomer units for a counterion between two points at any given moment of time. At large values of λ the charged groups seem to form small clusters surrounded by the neutral monomer units which restricts the movement of the counterions. This can be easily understood if one considers a copolymer with only a small fraction of charged groups; if these groups are distributed evenly (or randomly) and the system are homogeneous, the charged groups are essentially isolated from one another and the counterions form stable ion pairs with the monomer units. The diffusion is possible through the segmental movement and "hopping" of counterions between the ion pairs, but is it much slower than the diffusion in a microphase-separated state if all the charged groups form one block; in that case, the counterions can freely move within the domains formed by the ionic species.

**Conclusions**

In this work we investigated the phase behavior of melts of block-copolymers with one charged block by means of dissipative particle dynamics with explicit electrostatic interactions; such systems are a model of, for example, diblock-copolymers with one block being ionic liquid. We assumed that the blocks are fully compatible, i.e. all the Flory-Huggins χ parameters were equal to 0. This way, we showed that the charge correlation attraction solely can cause microphase separation with long-range order; a phase diagram was constructed by varying the volume fraction of the uncharged block and the electrostatic interaction parameter λ. Conventional phases like lamellae, hexagonally perforated lamellae and hexagonally packed cylinders were observed. The obtained phase diagram was compared to the phase diagram of corresponding neutral diblock-copolymers with non-zero χ-parameter between the beads of different blocks; one block of such copolymers consisted of "dumb-bell" monomer units to mimick the increase of the volume of each monomer due to the presense of counterions. Surprisingly, the differences between these phase diagrams are rather subtle; the same phases in the same order are observed, and the positions of the ODT points are similar if the λ-parameter is considered as an "effective" χ parameter. Therefore, the dimensionless parameters λ and χ seem to play the same role in the microphase separation phenomenon; the phase behavior seems to be almost independent of the exact nature of the driving force behind it.

Next, we investigated how the presence of free counterions and their entropy affects the behavior of the system. To that end, we studied the position of the ODT for lamellar structure depending on the chain length $N$. It turned out that while for the uncharged diblock-copolymer the product $\chi^{cr} N$ was almost independent of $N$, which is in line with the theoretical predictions for "classical" diblock-copolymers with symmetric blocks, for the diblock-copolymers with one charged block we observed a significant increase in $\lambda^{cr} N$ upon increasing $N$. We suppose it can be attributed to the fact that the counterion entropy prevents the formation of ordered structures, and its influence is more pronounced for longer chains since they undergo the transition to ordered structures at smaller values of $\lambda$, when the electrostatic energy becomes comparable to $k_b T$. This was supported by studying the ODT in diblock-copolymers with charged blocks and counterions cross-linked to the charged monomer units (thus forming dipole monomers). Indeed, the ODT for such systems was observed at significantly lower values of $\lambda$ with the difference being more pronounced at longer chain lengths $N$. Therefore, we can speculate that in the limit of $N \to \infty$ there will be a finite value of $\lambda$ necessary to observe the ODT in melts of diblock-copolymers with one charged block and free counterions unlike uncharged diblock-copolymers where $\chi^{cr} \to 0$ when $N \to \infty$.

Finally, we studied the diffusion of counterions in the obtained ordered structures and compared it to the case of a system with the same number of charged groups but homogeneous structure; the diffusion coefficient in the direction in the lamellar plane was found to be higher than in any direction in homogeneous structure. This can be attributed to the fact that in homogeneous structures the movement of the counterions is restricted by the presence of uncharged monomers surrounding the charged ones, while the formation of purely ionic phases in the case of ordered structures allow the counterions to move freely within the domains.

It is worth noting that in this work (as well as in the previous works [13] on the subject) the size of the monomer units and counterions is the same. However, the difference in the ion sizes can have a dramatic effect on the formation of dense ionic structures; for example, recent studies on the single polyelectrolyte chain collapse [25,28] have shown that the steric restrictions due to the ion size mismatch can significantly shift the position of the collapse. It seems that similar reasoning can be applied to the correlation attraction-driven microphase separation, and the latter is influenced by the ratio of the ion sizes.

Summarizing, in this work we shed some light on the complex and intriguing problem of self-organization of copolymers with charged units. The fact that the microphase separation is observed even at zero Flory-Huggins parameter can be used for creation of "high-$\chi$" copolymers:

the incorporation of charged groups (for example, ionic liquids) can significantly increase the segregation strength.

**Acknowledgments**

The financial support of the Russian Science Foundation (project 18-73-00128) is greatly acknowledged. The research is carried out using the equipment of the shared research facilities of HPC computing resources at Lomonosov Moscow State University.

**Notes**

The author declares no competing financial interest.